\documentclass{jfm}
\usepackage{epsfig}

\ifCUPmtlplainloaded \else
  \checkfont{eurm10}
  \iffontfound
    \IfFileExists{upmath.sty}
      {\typeout{^^JFound AMS Euler Roman fonts on the system,
                   using the 'upmath' package.^^J}%
       \usepackage{upmath}}
      {\typeout{^^JFound AMS Euler Roman fonts on the system, but you
                   dont seem to have the}%
       \typeout{'upmath' package installed. JFM.cls can take advantage
                 of these fonts,^^Jif you use 'upmath' package.^^J}%
      }
  \else
  \fi
\fi

\ifCUPmtlplainloaded \else
  \checkfont{msam10}
  \iffontfound
    \IfFileExists{amssymb.sty}
      {\typeout{^^JFound AMS Symbol fonts on the system, using the
                'amssymb' package.^^J}%
       \usepackage{amssymb}%
         \let\leq=\leqslant
         \let\geq=\geqslant
      }{}
  \fi
\fi

\ifCUPmtlplainloaded \else
  \IfFileExists{amsbsy.sty}
    {\typeout{^^JFound the 'amsbsy' package on the system, using it.^^J}%
     \usepackage{amsbsy}}
    {\providecommand\boldsymbol[1]{\mbox{\boldmath $##1$}}}
\fi

\newcommand\etb{\boldsymbol{\eta}}

\newsavebox{\astrutbox}
\sbox{\astrutbox}{\rule[-5pt]{0pt}{20pt}}

\def\g{\gamma}

\def\cd{\!\cdot\!}
\def\d{\delta}
\def\dd{\mbox{d}}
\def\ve{\varepsilon}

\def\o{\omega}
\def\r{{\bf r}}
\def\k{{\bf k}}
\def\q{{\bf q}}
\def\k{{\bf k}}

\def\v{{\bf v}}

\def\o{\omega}

\def\vp{\varphi}

\def\G{{\bf G}}

\def\R{{\bf R}}
\def\U{{\bf U}}

\newcommand{\gbout}[1]{} 

\begin{document}

\title[KH Instability with Periodic Plates]
{Kelvin-Helmholtz instabilities across periodic plates}


\author[Tom Chou]
{Tom Chou}

\affiliation{Dept. of Biomathematics and 
Dept. of Mathematics, UCLA, Los Angeles, CA 90095-1766 USA}

\maketitle

\begin{abstract}
We consider the linear stability of two inviscid fluids, in the
presence of gravity, sheared past each other and separated by an
flexible plate. Conditions for exponential growth of velocity
perturbations are found as functions of the flexural rigidity of the
plate and the shear rate. This Kelvin-Helmholtz instability is then
analysed in the presence of plates with spatially periodic (with
period $a$) flexural rigidity arising from, for example, a periodic
material variation.  The eigenvalues of this periodic system are
computed using Bloch's Theorem (Floquet Theory) that imposes specific
Fourier decompositions of the velocity potential and plate
deformations. We derive the nonhermitian matrix whose eigenvalues
determine the dispersion relation. Our dispersion relation shows
that plate periodicity generally destabilises the flow, compared to a
uniform plate with the same mean flexural rigidity. However, enhanced
destabilisation and stabilization can occur for disturbances with
wavelengths near an even multiple of the plate periodicity. The
sensitivity of flows with such wavelengths arises from the
nonpropagating, ``Bragg reflected'' modes coupled to the plate
periodicity through the boundary condition at the plate.
\end{abstract}

\section{Introduction}\label{sec:INTRO}

When two immiscible fluids are sheared relative to each other, the
flow, and the normally flat interface between them, may become
unstable.  This classical Kelvin-Helmholtz instability
(\cite{CHAND,DRAZIN,WHITHAM}) arises from perturbations of the
fluid velocity that grows into a vortex sheet due to local Bernoulli
variations in pressure.  The presence of gravity also tunes this
instability, depending on the mass density difference and shear rate 
between the two fluids.  If the denser fluid is in lower
gravitational potential, there is a critical shear rate beyond which
small perturbations grow exponentially. This instability first arises
in modes of the fluid velocity perturbation that have wavevectors
greater than a critical wavevector. All velocity perturbations with
sufficiently small wavelength become unstable as long as the shear
rate exceeds a critical value. However, very small wavelength
perturbations can be stabilised if additional restoring forces from,
say, an interfacial surface tension (\cite{WHITHAM}), or a separating
elastic plate are included. The Kelvin-Helmholtz instability, and its
generalisations, have been thoroughly studied, for example, in the
context of knotting in astrophysical jets (\cite{REES}), cloud top
entrainment instabilities (\cite{REISS}), magneto-hydrodynamic
instabilities (\cite{CHAND} and \cite{FRANK}), and the flutter of panels
(\cite{MILES}).

\begin{figure}
    \begin{center} 
        \includegraphics[height=2.0in]{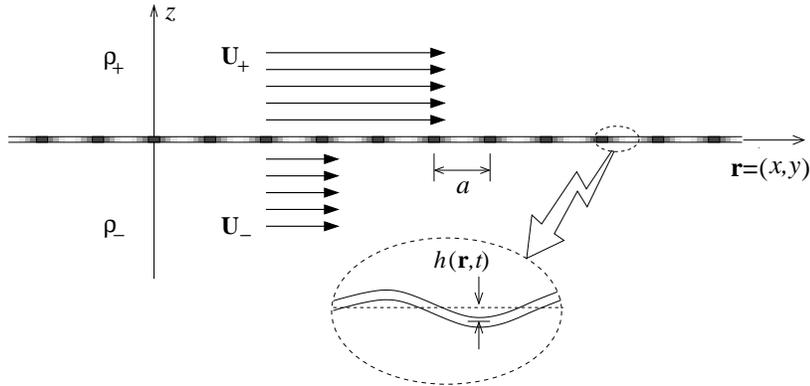}
      \end{center} 
\caption{Discontinuous shear flow separated by a periodically flexible
plate.}
\label{SLAB0}
\end{figure}

In the context of potential applications involving fluid-structure
interactions, we consider the Kelvin-Helmholtz instability in the
presence of an infinitesimally thin bendable plate separating the two
inviscid flows, as shown in Figure \ref{SLAB0}.  We assume the plate
is thin and inextensible, such that only bending modes can be excited.
Since elastic restoring forces stabilize short wavelength
perturbations, instabilities arise only for perturbations that have
wavevectors within a certain window of values. Furthermore, we
consider an elastic membrane that has spatially periodic properties,
such as mass density or bending rigidity, which might arise, for
example, in an engineered composite structure. By itself (in the
absence of fluid), a periodic plate supports stable bending waves that
propagate {\it except} at band gaps, {\it i.e.}  frequencies
corresponding to waves that have half-integer wavelengths in one
period of the structural variation (\cite{AM} and \cite{LIGHT}). Waves
near these wavelengths have vanishing group velocities and do not
propagate in an infinite periodic medium due to their multiple
``Bragg'' reflections and deconstructive interference.  Properties of
wave propagation through periodic media have been computed in numerous
physical settings, including acoustic waves moving through materials
with periodic sound speeds (\cite{SIGALAS}), electromagnetic
propagation through periodic dielectrics (\cite{LIGHT}), and water
wave propagation over periodic structures
(\cite{CHEN,CHOU,CHAN,HU,MEI1,PETER,PORTER}).

In this paper, we combine the periodic stabilizing membrane with an
otherwise unstable shear flow and study the complex dispersion
relation. A velocity perturbation with any wavevector, convected by
the base laminar flow forms a travelling wave. This travelling wave,
when coupled to the intrinsic periodicity of the elastic plate may be
Bragg reflected and not able to propagate within the region of
periodicity, altering its stability characteristics.
In order to explore this heuristic argument further, we formulate in
the next section the linear problem in terms of a vorticity-free
velocity perturbation and a small deformation of the separating
elastic membrane.  We review, in turn, a uniform plate subject to
destabilizing flow, and a free periodic plate that supports stable
waves.  The fully coupled problem, a periodic plate in destabilising
flow, is then presented in the Results section. We show that the
stability properties are affected most when the wavelengths of the
most unstable modes are close to the periodicity of the plate
flexibility.  Implications and extensions to related flow systems are
discussed the Discussion and Conclusions.

\section{Linear Model}\label{sec:LINEAR}

Consider the fluid-structure interaction depicted in Figure
\ref{SLAB0}.  Two incompressible fluids flowing uniformly in two
different directions, ${\bf U}_{+}$ and ${\bf U}_{-}$, in the $x-y$
plane are separated by a negligibly thin elastic plate.  Gravitational
acceleration is in the $-{\bf z}$ direction. Since viscosity will be
neglected, we will consider only potential flows.  The total velocity
in each region ($\pm$) is defined by the uniform flow $\U_{\pm}$ plus
a small irrotational perturbation $\v_{\pm} \equiv \nabla \vp_{\pm}$.
Upon imposing incompressibility, the upper and lower velocity
potentials, $\vp_{\pm}$, satisfy Laplace's equation:

\begin{equation}
(\nabla_{\perp}^{2} + \partial_{z}^{2})\vp_{\pm}(\r,z,t) = 0,\quad
\nabla_{\perp}^{2} = \partial_{x}^{2}+\partial_{y}^{2}.
\label{LAPLACE}
\end{equation}

Since $\partial_{z}\varphi\vert_{z=0} \approx v_{z}$ is the vertical
velocity of the surface, the kinematic conditions at the
plate-separated interface are

\begin{equation}
\partial_{t}h(\r,t) + \U_{\pm}\cdot\nabla_{\perp}h(\r,t) - 
\lim_{\ve\rightarrow 0}\partial_{z}\vp_{\pm}(\r,\pm\ve,t) = 0,
\label{KINEMATIC}
\end{equation}

\noindent where $h(\r,t)$ is the normal displacement of the plate.
The linearized normal stress balance at the interface,
including the normal forces arising from the flexible plate, is

\begin{equation}
\left[\rho\dot{\vp}+\rho\U\cd\nabla_{\perp}\vp+\rho gh(\r,t)\right]^{+}_{-}
=m(\r)\ddot{h}(\r,t) + \nabla_{\perp}^{2}(D(\r)\nabla_{\perp}^{2})h(\r,t),
\label{NSTRESS}
\end{equation}

\noindent where $\left[X\right]^{+}_{-} \equiv \lim_{\ve \rightarrow
0} X(z=\ve)-X(z=-\ve)$. In Eq. \ref{NSTRESS}, $m(\r)$ and $D(\r)$ are the
mass per unit area and the flexural rigidity of the plate,
respectively. Since $h(\r,t)$ and $\vp(\r,z,t)$ are defined
only up to a constant, we need only consider the nonzero wavevector
decompositions of  the plate displacement and velocity potential,

\begin{equation}
\displaystyle h(\r,t) = \sum_{\q\neq 0}\!\int\!h_{\q}(\omega)
e^{i\q\cdot\r-i\omega t}{\dd \o \over 2\pi} \quad \mbox{and} \quad
\vp_{\pm}(\r,z,t) = \sum_{\q\neq 0}\!\displaystyle
\int\!\varphi^{\pm}_{\q}(\omega) e^{i\q\cdot\r-i\omega t
\mp\vert\G-\q\vert z}{\dd \o \over 2\pi}.
\label{EXPANSIONS}
\end{equation}



\noindent In Eqs. \ref{EXPANSIONS}, we have also assumed a continuous
time-harmonic decomposition.  The representation of $\vp_{\pm}(\r, z,
t)$ automatically satisfies Eq. \ref{LAPLACE} and the boundary
conditions at $z\rightarrow \pm\infty$.

The periodic plate mass density and rigidity functions that obey
$m(\r)= m(\r+\R)$ and $D(\r)=D(\r+\R)$, where $\R=n_{1}a_{1}{\bf
e}_{1} + n_{2}a_{2}{\bf e}_{2}\,$ ($n_{1},n_{2}$ integers) is any
basis vector describing the periodicity. Here, $a_{1}, a_{2}$ are the
periodicities of $m(\r)$ and $D(\r)$ in the ${\bf e}_{1}, {\bf e}_{2}$
directions.  The periodic functions $m(\r)$ and $D(\r)$ can thus be
decomposed into a Fourier sum over reciprocal lattice vectors $\G$,

\begin{equation}
m(\r)=\sum_{G}m(\G)e^{i\G\cdot\r} \,\quad \mbox{and}\quad\,   
D(\r) = \sum_{G}D(\G)e^{i\G\cdot\r}.
\label{DECOMP}
\end{equation}

\noindent The reciprocal lattice vectors $\G = n_{1}\G_{1} + n_{2}\G_{2}$
are defined by the reciprocal lattice basis vectors

\begin{equation}
\G_{1} = {2\pi\over a_{1}} {{\bf e}_{2}\times{\bf z}\over
{\bf e}_{1}\cdot({\bf e}_{2}\times {\bf z})} \,\quad\mbox{and}\quad\,
\G_{2} = {2\pi\over a_{2}} {{\bf z}\times{\bf e}_{1}\over
{\bf e}_{1}\cdot({\bf e}_{2}\times {\bf z})}.
\end{equation}

Finally, we assume a large plate with dimensions $L_{1}\times L_{2}$
that fit an integer multiple of the period ($L_{1} = N_{1}a_{1}$ and
$L_{2} = N_{2}a_{2}$) and impose periodic boundary
conditions on the entire system: $m(\r + N_{i}a_{i}{\bf e}_{i}) =
m(\r)$ and $D(\r + N_{i}a_{i}{\bf e}_{i}) = D(\r)$ for $i=1,2$.  Under
this construction, the wavevector sum in the decompositions
\ref{EXPANSIONS} are taken over $\q = \sum_{n_{i}\neq
0}(n_{i}/N_{i})\G_{i}$. In the $L_{1},L_{2}\rightarrow \infty$ limit,
sums over $\q$ become the Cauchy principle value of the corresponding
Riemann integral.

After substitution of Eqs. \ref{EXPANSIONS} into the 
kinematic conditions \ref{KINEMATIC}, we find 
the relationship

\begin{equation}
\varphi^{\pm}_{\q}(\o) = \pm {i(\o - \U_{\pm}\cd\q)\over \vert
\q\vert}h_{\q}(\o), \qquad \q\neq 0.
\label{PHIH}
\end{equation}

\noindent Upon substitution of Eqs. \ref{EXPANSIONS}, \ref{DECOMP},
and \ref{PHIH} into Eq. \ref{NSTRESS}, and exploiting the
orthogonality of the $e^{i\q\cdot\r}$ basis functions,

\begin{equation}
\begin{array}{l}
\left[(\rho_{+}+\rho_{-})\o^{2} - 
2\o (\rho_{+}\U_{+} + \rho_{-}\U_{-})\cd\q + 
\rho_{+}(\q\cd\U_{+})^{2}+\rho_{-}(\q\cd\U_{-})^{2} + 
(\rho_{+}-\rho_{-})g\vert\q\vert\right]h_{\q} \\[13pt]
\:\hspace{2cm} = 
-\o^{2}\vert\q\vert\sum_{\G'}m(\G')h_{\q-\G'} + 
\sum_{\G'}\vert\q\vert^{3}\vert\q-\G'\vert^{2}D(\G')h_{\q-\G'}.
\end{array}
\label{INTEGRAL}
\end{equation}

\noindent Equation \ref{INTEGRAL} shows that each coefficient $h_{\q}$
is linked with $N_{1}\times N_{2}$ other coefficients
$h_{\q-\G'}$. This problem can be solved numerically by truncating
$h_{\q-\G} \approx 0$ for large $\vert\G\vert$ where $m(\G)$ and
$D(\G)$ are small.  Alternatively, Eq. \ref{INTEGRAL} can be organized
according to the ``reduced zone scheme,'' by defining $\k = \q-\G$,
where $\G$ is such that $\k$ is restricted in the first Brillouin zone
defined by $\vert\k\cdot{\bf e}_{i}\vert < \vert\G_{i}\cdot{\bf
e}_{i}\vert/2$ (\cite{AM}).  In this representation, we need only
consider the $N_{1}\times N_{2}$ eigenvalues, each with index $\k$
inside the first Brillouin zone. The problem is thus cast into a
quadratic eigenvalue problem defined by

\begin{equation}
\sum_{\G'}\left[A_{\k}(\G,\G')\omega^{2} + B_{\k}(\G,\G')\omega +
C_{\k}(\G,\G')\right]h_{\k-\G'}(\o) = 0,
\label{MATRIX2}
\end{equation}

\noindent where 

\begin{equation}
\begin{array}{l}
\displaystyle A_{\k}(\G,\G') = {\rho_{+}+\rho_{-} \over
\vert\k-\G\vert}\d_{\G,\G'} + m(\G'-\G) \\[13pt] \displaystyle
B_{\k}(\G,\G') = - 2\left[\rho_{+}{\U_{+}\cd(\k-\G)\over \vert\k-\G\vert} +
\rho_{-}{\U_{-}\cd(\k-\G)\over
\vert\k-\G\vert}\right]\d_{\G,\G'}\\[13pt] \displaystyle C_{\k}(\G,\G') =
\left[\rho_{+}{(\U_{+}\cd(\k-\G))^{2}\over \vert\k-\G\vert}+
\rho_{-}{(\U_{-}\cd(\k-\G))^{2}\over
\vert\k-\G\vert}+(\rho_{+}-\rho_{-})g\right]\d_{\G,\G'} \\[13pt]
\:\displaystyle
\hspace{6cm} -\vert\k-\G\vert^{2}\vert\k-\G'\vert^{2}D(\G'-\G).
\end{array}
\end{equation}

\noindent Equation \ref{MATRIX2} is a matrix equation for
$h_{\k-\G}(\o)$ that has solutions for only certain 
$\o(\k)$. 
The full dispersion relation can be recovered by assigning the
appropriate translation $\G$ in the eigenvalues $\o(\k-\G)$ to
$\omega(\q)$ for wavevectors $\q$ outside the first Brillouin zone.
The system becomes unstable when $\omega(\q)$ acquires an imaginary
component.

In anticipation of a length scale $a$ associated with the periodicity
in the plate properties, we henceforth nondimensionalise all
parameters and eliminate $\rho_{-}, g$ and $a$ according to

\begin{equation}
\begin{array}{l}
\displaystyle \r \rightarrow \r/a, \, \, \G \rightarrow a\G,\,\,
\k\rightarrow a\k,\,\, \o \rightarrow \sqrt{{a\over g}}\o, \,\, \U_{\pm} \rightarrow {\U_{\pm}\over \sqrt{ag}} \\[13pt]
\displaystyle D\rightarrow {D \over a^{4}\rho_{-}g},\,\, m \rightarrow {m\over
a\rho_{-}},\,\,\,\mbox{and}\,\,\,  h_{\k}(\G) \rightarrow h_{\k}(\G)/a.
\end{array}
\end{equation}



\subsection{A Uniform Plate in the Presence of Flow}

First consider the standard Kelvin-Helmholtz instability in the
presence of a uniform elastic plate, where $m=m_{0}$ and $D(\r) =
D_{0}$ are constants.  \cite{MILES01} treated a similar problem of a
thin boundary layer over an elastic plate. Since $D(\G) =
D_{0}\delta_{\G,0}$ and $m(\G) = m_{0}\delta_{\G,0}$,
Eq. \ref{MATRIX2} is diagonal and is satisfied when

\begin{equation}
\Gamma(q)\omega_{\pm}(\q) = (\g\U_{+}+\U_{-})\cd\q \pm
\sqrt{\Delta(\q)},
\label{OMEGA}
\end{equation}

\noindent where $\Gamma(q)\equiv \g+1+m_{0}q$, and 
the discriminant

\begin{equation}
\Delta(\q) \equiv
\Gamma(q)(D_{0}q^{5}+(1-\g)q)-\g((\U_{+}-\U_{-})\cd\q)^{2}-m_{0}q
(\g(\U_{+}\cd\q)^{2}+(\U_{-}\cd\q)^{2}).
\label{DELTA}
\end{equation}

\noindent The eigenvalues $\o_{\pm}$ become complex when $\Delta(\q)$
becomes negative, leading to exponentially growing perturbations. In
the limit where the influence of the plate is negligible ($m_{0},
D_{0}\approx 0$) the dispersion relation reduces to that of the
standard Kelvin-Helmholtz instability. The inclusion of membrane
stiffness ($D_{0}> 0$) tends to stabilise small wavelength modes.  For
any given $m_{0}\geq 0$, the pairs of critical $D_{0}$ and $U$
(denoted $D_{*}$ and $U_{*}$) that give the onset of instability at a
chosen wavevector $q_{*}$ can be expressed explicitly in two flow
configurations where $\U_{\pm}$ and the $\q$ corresponding to the most
unstable velocity perturbations are all collinear.  When $\U_{\pm}$ are
collinear, the onset of instability arises when $\Delta(\q) = 0$ is
solved by two critical repeated, real roots, $q_{*}$, in addition to
the root at $q=0$ and two complex conjugate roots.  For a static
configuration of the lower fluid ($\U_{-}=0$ in the reference frame of
the plate), we find

\begin{equation}
\begin{array}{rcl}
\displaystyle D_{*} &=& \displaystyle {(1+m_{0}q_{*})^2 -
\g^{2}(1+2m_{0}q_{*})-\g m_{0}^{2}q_{*}^{2}\over 
q_{*}^{4}(3(1+\g+m_{0}^{2}q_{*}^{2})+2(\g+3)m_{0}q_{*})}\\[13pt]
\displaystyle U_{*}^{2} &=& \displaystyle
{4(1-\g)(1+\g+m_{0}q_{*})^{2} \over 
\g q_{*}(3(1+\g+m_{0}^{2}q_{*}^{2})+2(\g+3)m_{0}q_{*})}.
\end{array}
\label{DUSTAR}
\end{equation}


\begin{figure}[h!]
    \begin{center} 
        \includegraphics[height=2.2in]{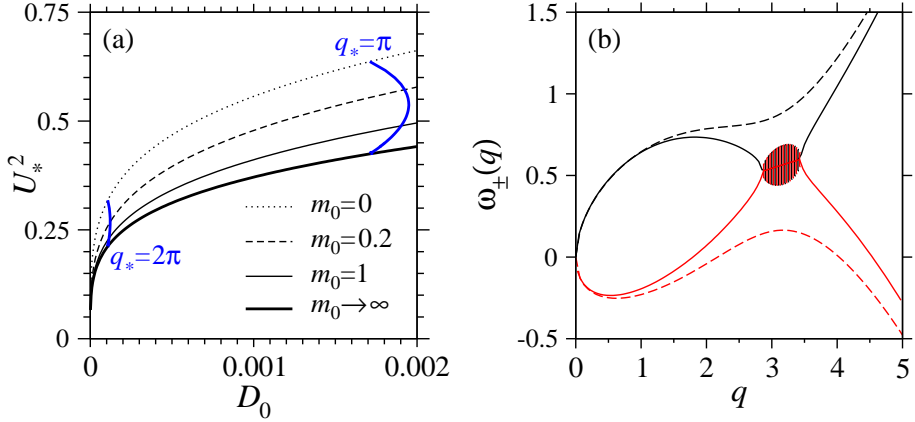}
      \end{center} 
\caption{(a) The curve relating the critical value of the shear
$U_{*}$ to the constant flexural rigidity $D_{0}$, for various values
of the uniform plate mass density $m_{0}$. For  velocities
greater than $U_{*}$, the flow becomes unstable. For different plate
mass densities, the sets of $D_{*}, U_{*}$ that give rise to an
initial instability at wavevectors $q_{*}=\pi$ and $q_{*}=2\pi$ are
shown by the black arcs spanning the curves. (b) The dispersion
relation for a uniform plate just after the onset of instability.}
\label{DU}
\end{figure}

\noindent Alternatively, for a chosen value of $D_{0}$, the critical
$U_{*}$ and the corresponding initially unstable wavevector can be
found implicitly.  The critical velocities $U_{*}$, as a function of
$D_{0}$ and $m_{0}$ are shown in Fig. \ref{DU}(a).  Note that the
critical velocity $U_{*}$ {\it decreases} as the plate mass density
$m_{0}$ increases. The limit $m_{0}\rightarrow \infty$ is different
from the $\rho_{\pm}\rightarrow 0$ limit treated in the next section.
Here, the plate flexibility remains constant as its mass density
increases.  Although the magnitude of $\o_{\pm}(q)$ decreases, the
discriminant (Eq. \ref{DELTA}) becomes negative at smaller $U_{*}$ due
to the increased inertial forces of he plate back on the fluid.

The two branches of the full dispersion relation $\omega_{\pm}(q)$ are
plotted in Fig. \ref{DU}(b).  We have chosen parameters ($D_{*}(\pi)$
and $U_{*}^{2}$ defined by Eqs. \ref{DUSTAR}) such that the onset of
instability arises at $q_{*}\approx \pi$. For $U = 1/\sqrt{2} <
U_{*}(q_{*}=\pi) = 0.756320$, $\o_{\pm}(q)$ are real, as shown by the dashed
curves.  We also plot (solid curves) the dispersion relation for $U =
0.761577 > U_{*}(q_{*}=\pi) = 0.75632$, whereupon a window of instability
opens near $q_{*}=\pi$. The magnitude of the complex parts of
$\o_{\pm}(q)$ are indicated by the height of the shaded region.

In the second flow configuration, we assume a vanishing net momentum in the 
reference frame of the plate: $\U_{-}=-\gamma \U_{+}$. In this case, we find
independent of $m_{0}$,

\begin{equation}
D_{*} = {1-\g \over 3q_{*}^{4}}\quad \mbox{and} \quad U_{*}^{2}
={4(1-\g) \over 3\g(\g+1)q_{*}},
\label{DUSTAR2}
\end{equation}

\noindent and the symmetry $\omega_{+}(q) = -\omega_{-}(q)$.  For
$D_{0} \lesssim D_{*}$ and/or $U\gtrsim U_{*}$, as defined by
Eq. \ref{DUSTAR} or Eq. \ref{DUSTAR2}, $\omega(q\approx q_{*})$
becomes complex and instability arises for a small window of
wavevectors near $q_{*}$.


\subsection{Dispersion Relation for a Free Periodic Plate}\label{subsec:FP}

Now consider the limit of an isolated (but periodically structured)
plate where the inertia of the bounding fluid is negligible. If the
plate were uniform, the dispersion relation in the limit
$\rho_{\pm}\rightarrow 0$ is simply $m_{0}\o^{2} = D_{0}q^{4}$. For a
periodically structured plate ($m(\G\neq 0)\neq 0, D(\G\neq 0) \neq
0$), the dispersion relation can be found from Eq. \ref{MATRIX2} with
the simplified matrices $A(\G-\G') = m(\G-\G'), B(\G-\G')=0$, and
$C(\G-\G') = -\vert\k+\G\vert^{2}\vert\k+\G'\vert^{2}D(\G-\G')$.
Henceforth, we assume the periodicities in the plate to be sinusoidal
in the ${\bf x}$-direction and take the forms

\begin{equation}
m(\r) = m_{0} + 2m_{1}\cos 2\pi x; \quad D(\r) = D_{0} + 2D_{1}\cos 2\pi x,
\end{equation}

Although the mean mass density and bending rigidity are unchanged from
$m_{0}$ and $D_{0}$, respectively, the variations $m_{1}$ and/or
$D_{1}$ will change the dispersion relation. Specifically, periodic
variations will break the dispersion relation at wavevectors
corresponding to waves that cannot propagate through an infinite,
periodic medium. At each of these wavevectors (the ``Bragg planes''),
the waves (half-integers of which fit into a period) become
standing. Corresponding to these nonpropagating waves are gaps in the
frequency (band gaps, or stop bands) within which an infinitely
periodic material cannot be excited.

\begin{figure}[h!]
    \begin{center} 
        \includegraphics[height=2.2in]{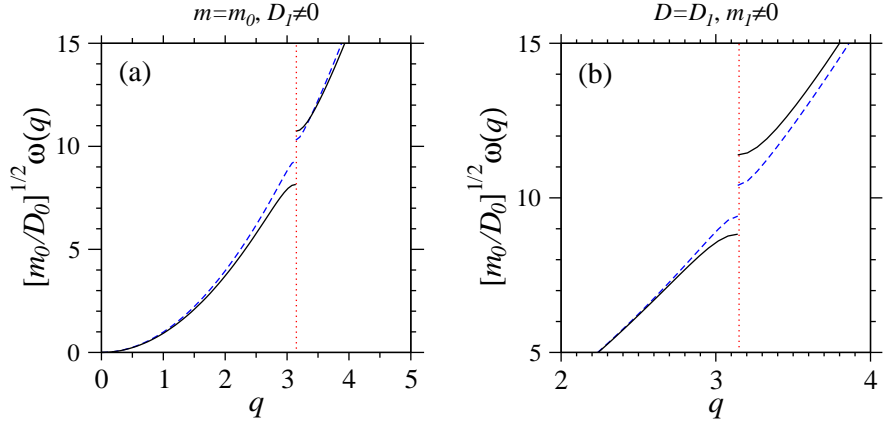}
      \end{center} 
\caption{(a) The normalized dispersion relation
$\sqrt{m_{0}/D_{0}}\,\o_{\pm}(q)$ for $m_{1}=0$, $D_{1}/D_{0} = 0.1$
(dashed curve) and $D_{1}/D_{0}=0.25$ (solid curve).  (b) A blow-up of
the dispersion relation for $D_{1}=0, m_{1}/m_{0}=0.1$ (dashed curve)
and $m_{1}/m_{0}=0.25$ (solid curve).}
\label{DISPBG}
\end{figure}

Figures \ref{DISPBG}a,b show the dispersion relation of an isolated
plate with periodicities in the bending rigidity (a), and mass density
(b).  The frequency gaps increase as the contrast $m_{1}$ or $D_{1}$
is increased. From consideration of the eigenvectors, the standing
modes near the band gaps can be shown to have either nodes or
antinodes over the high flexural rigidity (or high mass density)
regions, depending on which side of the band gap the wavevector $q$
lies. These behaviour generally arise in wave propagation through
periodic media.  When the plate is coupled to flow, it will be the
near equal-wavelength modes immediately straddling the band gaps, with
different frequencies and $\sim 90^{\circ}$ out-of-phase, that will
qualitatively affect the stability near the band gaps.

\section{Results for the Coupled Problem}\label{sec:FCP}

We now combine the known results of the previous sections and consider
the fully coupled problem where fluid instabilities and Bragg
reflection interact. For simplicity, we consider the plate periodicity
wavevectors ($\G = 2\pi n/a {\bf x}$) and the uniform flows ${\bf
U}_{\pm}=U_{\pm}{\bf x}$ to be aligned. In the full problem, the
dispersion relation $\o(q)$ may be both complex (signalling regions of
instability) and discontinuous at band gaps where
$(\partial\omega/\partial q) = 0$.  The full dispersion relation is
found from solving the quadratic eigenvalue problem given by
Eq. \ref{MATRIX2}.  This problem can be expressed in linear eigenvalue
form,

\begin{equation}
({\bf M}-\omega {\bf I})\cdot\etb = 0,
\label{LINEAREIGEN}
\end{equation}

\noindent with $\etb = (h_{\k}, \o h_{\k})^{T}$, and 

\begin{equation}
{\bf M} = \left(\begin{array}{cc}
{\bf 0} & {\bf I} \\[13pt]
-{\bf A}^{-1}{\bf C} & -{\bf A}^{-1}{\bf B} \end{array}\right).
\label{ACBMATRIX}
\end{equation}

The dispersion relation $\o(\k)$ is found from the eigenvalues of the
nonhermitian matrix ${\bf M}$. For the parameters explored, we find
numerical convergence of the lowest handful of eigenvalues with 40 or
fewer modes $\G, \G'$. Therefore, we truncate the system at
approximately 60 modes (where ${\bf M}$ is a $120\times 120$
matrix). This is more than sufficient to obtain numerical accuracy for
the lower eigenvalues at all relevant wavevectors.  The eigenvalues of
${\bf M}$ are sorted from the numerical solutions and replotted in the
extended zone scheme where the wavevector $0\leq \vert\q\vert <
\infty$.  When unfolding the eigenvalues $\omega(\k)$, we make use of
the fact the functions $\omega(0\leq \vert\q\vert < \infty)$ in the
normal extended zone scheme come in pairs $\omega_{\pm}(\q)$ and have
the symmetry property $\omega_{\pm}(-\vert\q\vert) =
-\o_{\mp}(\vert\q\vert)$.  This property, along with the consideration
of the corresponding eigenvectors at each wavevector $k$ allows us to
reconstruct $\omega_{\pm}(q)$ from $\omega_{\pm}(k)$.

\begin{figure}
    \begin{center} 
        \includegraphics[height=4.0in]{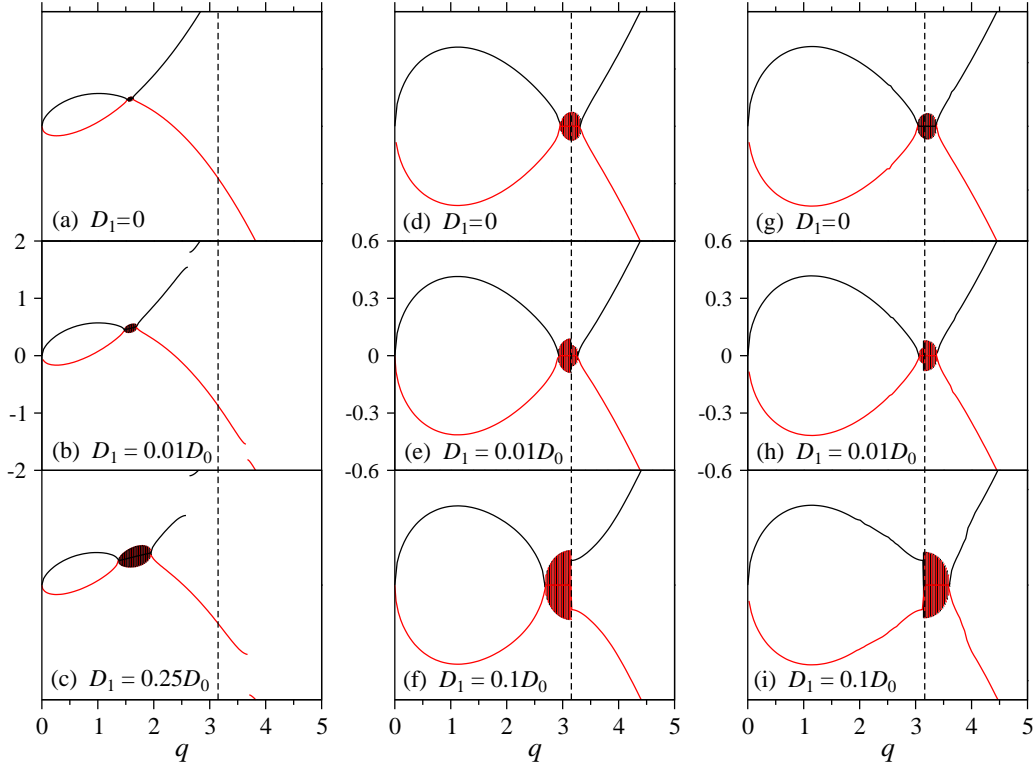}
      \end{center} 
\caption{(a-c) The complex dispersion relation for a system just above
the instability threshold.  The instability is first triggered at
$q_{*} = \pi/2$ (which arises for parameters $\gamma =1/2,
m_{0}=0.2, D_{0} \leq D_{*} = 0.0298349$, and $U^{2} \geq U_{*}^{2} =
1.19809$). As $D_{1}$ increases, the instability bubble
(Im$\{\o_{\pm}\}$) increases in size as more wavevectors become
unstable. (d-f) Dispersion relation for instabilities that first arise
at $q_{*}=3.14\lesssim \pi$. As $D_{1}$ is increased, the modes with $q < \pi$ are
destabilised, while those with $q>\pi$ are stabilised. (g-i)
Increasing $D_{1}$ when the incipient instability starts at
$q_{*}=3.25 >\pi$ destabilises modes with $q> \pi$ and stabilises
those with $q< \pi$.}
\label{FULL}
\end{figure}

In Figures \ref{FULL}(a-c) we show the effects of increasing the
strength of the periodic rigidity $D_{1}/D_{0}$ for the case
$\U_{-}=0$. In Fig. \ref{FULL}(a) we show a system with an incipient
instability at $q \approx q_{*} = \pi/2$.  As $D_{1}$ is increased
(Figs. \ref{FULL}(b-c)) the bubble of instability grows modestly,
while gaps in the dispersion relation also appear. The destabilising
effect of rigidities with periodicity $a \gg 2\pi/q_{*}$ can be
understood in terms of a slight effective softening of the plate to
bending modes with wavelengths greater than $a$\footnote{The effective
long wavelength bending rigidity of a plate made from alternating
equal strips of two materials with individual rigidities
$D_{0}\pm\Delta D$ is $D_{eff} = 2/(1/(D_{0}+\Delta D)+1/(D_{0}-\Delta
D)) \approx D_{0} - (\Delta D)^2/D_{0}$}.

Note that the gaps here do not appear at the Brillouin zone edge
$q=\pi$ defined by a system's static periodic structure as in the free
plate problem (Figs. \ref{DISPBG}) and in most other settings.
Rather, they appear symmetrically around the band edge due to a
doppler shift arising from the nonvanishing net momentum of the
system. Gaps occurring away from the zone boundaries have been shown
by \cite{BERMEL} to arise in electromagnetic propagation through
deformed cholesteric elastomers.

Now consider an incipient instability that arises at higher
wavevectors, near the first band gap at $q=\pi$. Assume for
simplicity, the zero net momentum ($\U_{-}=-\gamma \U_{+}$) case where
${\bf B}_{\k}(\G,\G') = 0$ in Eq. \ref{MATRIX2} and
$\omega_{\pm}(q)=-\o_{\mp}(q)$. Figures \ref{FULL}(d-f) show the
effects of increasing $D_{1}$ when the initial instability starts at
$q_{*}=3.14 \lesssim \pi$.  As the plate heterogeneity is increased,
modes with wavevector $q < \pi$ are destabilised, while those with
$q>\pi$ are stabilised. Conversely, if the incipient instability
arises at $q_{*} \gtrsim \pi$, as in Figs. \ref{FULL}(g-i) (where
$q_{*}=3.25$), increasing $D_{1}$ will destabilise modes with $q>\pi$,
while stabilising those with $q<\pi$. The discontinuous change in the
eigenvalues across the Bragg plane $q=\pi$ occurs first in the
imaginary component and the growth rate $\mbox{Im}\{\o(q)\}$ is
discontinuous. As the contrast $D_{1}/D_{0}$ is further increased,
$\mbox{Re}\{\o(q)\}$ also becomes discontinuous. This behavior can be
explicitly seen with perturbation analysis of Eq. \ref{ACBMATRIX} by
truncating the matrix and finding the four eigenvalues as $q
\rightarrow \pi$.

\section{Discussion and Conclusions}

We have analysed the properties of the instabilities of two fluids
uniformly flowing past each other, separated by an elastic plate with
periodic bending rigidity and/or mass density. Our generalization of
the Kelvin-Helmholtz instability can be quantified by finding the
eigenvalues and eigenvectors of a nonhermitian matrix
(Eq. \ref{MATRIX2}).  The imaginary components of the eigenvalues
$\o(q)$ determine the rate of unstable growth for eigenmodes of
wavevector $q$. We find that an imposed plate periodicity generally
destabilises the flow but that the destabilisation mainly occurs in
the band of wavevectors on the side of the initially most unstable
mode. Modes with wavevectors on the other side of the band gap from
the most unstable wavevector $q_{*}$ are stabilised as $D_{1}$ is
increased.

The destabilisation/stabilisation that occurs on either side of the
band gap can be understood in terms of $\sim 90^{\circ}$ out-of-phase
standing modes with wavevectors $q=(2n-1)\pi \pm \varepsilon$
immediately straddling the $n^{th}$ band gap. For example, if $q_{*}
\gtrsim (2n-1)\pi$, increasing $D_{1}$ destabilises the waves with $q
\approx q_{*}$ since their antinodes occur predominantly over regions
of lower rigidity. Modes with $q < (2n-1)\pi$, on the other side of
the $n^{th}$ band gap, can be significantly stabilised since they are
approximately $90^{\circ}$ out-of-phase, and their antinodes sample a
stiffer plate. The same argument applies for $q_{*} \lesssim
(2n-1)\pi$.

Our results suggest possible stability control strategies.  By adding
restoring forces (such as bending rigidity) with periodicity near the
wavelength of the most unstable mode, stability can be enhanced or
diminished at wavelengths either less than or greater than the band
gap wavelength. Stabilisation of specific modes against
Kelvin-Helmholtz instabilities can be achieved, but at the expense of
destabilisation of other nearby modes.

Extensions of our approach to stability analysis in other
fluid-structure systems should be possible. For example, the
Orr-Sommerfeld problem (\cite{ORSZAG71}) in nonuniform Poiseuille flow
through periodically elastic pipes can be similarly solved.  A
physical realization of flow through elastically periodic pipes may
have arisen in the experiments of \cite{KRAMER} in which damping
elements were periodically embedded along an elastic pipe.  The
Kelvin-Helmholtz type instability has been only briefly mentioned in
this context by \cite{BENJAMIN} and \cite{LANDAHL}, and only in
uniform pipes.  Moreover, the effects of periodicity on other classes
of instabilities ({\it e.g.} Tollmien-Schlichting) have not been
explored. It would also be interesting to consider the effects of
dissipation, in both the fluid flow and the viscoelastic plate, on the
stability properties of a periodic system.

\acknowledgements

The author was supported by the US National Science Foundation through
grant DMS-0349195, and the US National Institutes of Health via grant
K25AI058672.

\end{document}